\documentclass[letterpaper]{article} 
\usepackage{aaai23}  
\usepackage{times}  
\usepackage{helvet}  
\usepackage{courier}  
\usepackage[hyphens]{url}  
\usepackage{graphicx} 
\usepackage{natbib}  
\usepackage{caption} 
\usepackage{algorithm}
\usepackage{algorithmic}
\usepackage{newfloat}
\usepackage{listings}
\DeclareCaptionStyle{ruled}{labelfont=normalfont,labelsep=colon,strut=off} 
\floatstyle{ruled}
\newfloat{listing}{tb}{lst}{}
\floatname{listing}{Listing}
\usepackage{multirow}
\usepackage{booktabs}
\usepackage[hang]{subfigure}
\usepackage{tikz}
\usepackage{arydshln}

\title{Intuitive Access to Smartphone Settings Using Relevance Model Trained by Contrastive Learning}

\author{
    Joonyoung Kim, Kangwook Lee, Haebin Shin, Hurnjoo Lee,\\ Sechun Kang, Byunguk Choi, Dong Shin, Joohyung Lee
}
\affiliations{
    Samsung Research\\

    56, Seongchon-gil, Seocho-gu, Seoul, Republic of Korea\\
    \{joon0.kim, kw.brian.lee, haebin0.shin, hurnjoo.lee, sechun.kang, byunguk.choi, d0104.shin\}@samsung.com, 
    joolee@asu.edu
}

\begin{document}
\maketitle
\begin{abstract}
The more new features that are being added to smartphones, the harder it becomes for users to find them. This is because the feature names are usually short, and there are just too many to remember. In such a case, the users may want to ask contextual queries that describe the features they are looking for, but the standard term frequency-based search cannot process them. This paper presents a novel retrieval system for mobile features that accepts intuitive and contextual search queries. We trained a relevance model via contrastive learning from a pre-trained language model to perceive the contextual relevance between query embeddings and indexed mobile features. Also, to make it run efficiently on-device using minimal resources, we applied knowledge distillation to compress the model without degrading much performance. To verify the feasibility of our method, we collected test queries and conducted comparative experiments with the currently deployed search baselines. The results show that our system outperforms the others on contextual sentence queries and even on usual keyword-based queries.
\end{abstract}

\section{Introduction}
Every new smartphone release is accompanied by many new features to attract users. Ironically, it becomes harder for the users to access them because the feature names in Settings are usually concise, each manufacturer calls them differently, and there are too many of them for the users to remember the exact terms. This differs from the standard document search, where the target document is long enough to contain words that capture the users' intent. 

For finding a menu in Settings, users may want to ask contextual queries that describe the features they are looking for, but the traditional keyword-based search engines such as TF-IDF \cite{salton1986introduction} and BM25 \cite{robertson1995okapi} cannot process them. To overcome the limitation of handling the diversity of users' contextual queries, the current search engines in Android mobiles use look-up tables, but capturing all variations of users' utterances into the look-up tables is not a scalable solution. Also, it is hard to maintain such tables. The problem becomes more severe with the new release with more features. 

This paper presents a novel retrieval system for mobile features that accepts intuitive and contextual search queries. 
The proposed approach can distinguish the relevance among target candidates by understanding contextual semantics via the relevance model trained in a contrastive manner.
Furthermore, we applied knowledge distillation to make it run on-device using minimal resources \cite{hinton2015distilling}. 
By transferring the knowledge in the large model to a compact model with less layers and smaller hidden dimensions, we could reduce the model size to about 1/5 of the original one with only 5\%  performance degradation.

To verify our method's feasibility and efficiency, we conducted comparative experiments with the currently deployed keyword-based search systems such as OneUI 3.1 and iOS 15.6, which shows that our retrieval system performs better not only on relaxed keyword queries, but also on keyword queries by handling synonyms and compound nouns well. 

In summary, this paper makes the following contributions. 
\begin{itemize}
	\item We propose the contextual retrieval system for mobile features using a relevance model trained in a contrastive manner.
	\item To deploy our method on-device, we successfully applied knowledge distillation to reduce the model size without degrading much performance.
	\item We demonstrate the advantages and robustness of our system through comparative experiments on various types of queries.
\end{itemize}

\begin{figure*}[t]
\centering
\includegraphics[width=2.0\columnwidth]{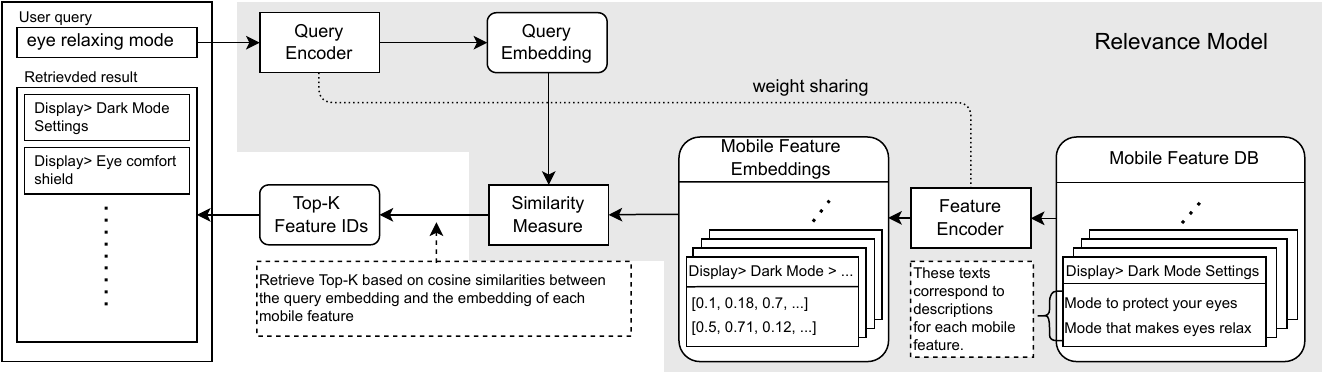}
\caption{The overview of the retrieval system for mobile features. When a user enters a query, it computes the query embedding through a query encoder. Then, it retrieves top-K relevant candidates based on the cosine similarities between the query embedding and pre-computed embeddings of candidate features via the relevance model.}
\label{fig1}
\end{figure*}

\section{Related Works}

\subsection{Keyword Search}

The classical keyword-based search methods, such as TF-IDF \cite{salton1986introduction} and BM25 \cite{robertson1995okapi}, depend on the exact match between terms in a query and indexed documents while considering Inverse Document Frequency \cite{luhn1957statistical, jones1972statistical}.
However, the more features that are being added to a smartphone, the harder it becomes to retrieve relevant search results: the retrieval quality of keyword-based search highly depends on a user's prior knowledge of a target domain to create relevant queries. 


\subsection{Neural Search}

The neural search engine, which works with vector representations containing textual meanings, can retrieve semantically similar documents even if none of the query terms are matched lexically \cite{DBLP:journals/corr/MitraC17, Mitra2018AnIT, 10.1145/3038912.3052579, 10.1145/3077136.3080809, 10.1145/3077136.3080832}.
In \cite{horita2019enhancing}, the authors utilized Word2Vec \cite{10.5555/2999792.2999959} to support the search on the features in the Settings app with their semantic embeddings.
However, the approach is vulnerable to out-of-vocabulary and cannot consider the contextual semantics of a query.
Two types of neural search architectures that utilize pre-trained language models, such as BERT \cite{devlin2019bert}, RoBERTa \cite{liu2019roberta}, and ELECTRA \cite{clark2020electra}, have been proposed to overcome the limitation of the fixed embedding-based approaches.  Cross-encoder \cite{wolf2019transfertransfo, vig2019comparison} shows good accuracy but has a high computational complexity in computing the embeddings of the query-indexed data combinations each time a new query enters.
On the other hand, Bi-encoder \cite{mazare2018training, dinan2018wizard, reimers2019sentence} seeks efficiency with little accuracy degradation by measuring the relevance of a query embedding over the pre-computed embeddings of the indexed data.

\subsection{Knowledge Distillation}
One effective way to compress a large model is knowledge distillation (KD). It exploits a large model's logits as soft labels for a student model. In a pioneering work, \cite{hinton2015distilling} proposes this  mechanism assuming output logits represent the knowledge of neural networks. Subsequent studies on KD have been proposed: Knowledge types \cite{kim2016sequence, turc2019well, park2019relational}, distillation algorithms \cite{zhang2019your}, online KD \cite{anil2018large}, and a theoretical explanation \cite{phuong2019towards}.

\section{Effective Retrieval}
Here we propose a retrieval system to ease users in finding a desired feature.
It has a bi-encoder architecture utilizing pre-computed embeddings of indexed data for computational efficiency.
To train an encoder, we design a Siamese network \cite{koch2015siamese} and train it in a contrastive learning manner with refined relevant query-document pairs.
To exploit the prior knowledge of language understanding, pre-trained language models are used for initialization. Specifically, we follow the training strategy of RoBERTa \cite{liu2019roberta}.

As Figure \ref{fig1} describes, in an offline manner, the database of mobile features is pre-computed as feature embeddings through the feature encoder to reduce inference latency. When a user enters a query, the query encoder turns it into a query embedding. Then the relevance model estimates the cosine similarities (\textit{relevance}) between the query embedding and the target features. Based on these relevance scores, it retrieves Top-K mobile features for the query. The system is able to capture the contextual meaning by exploiting an expressively abundant representation of a language model.

\subsection{Pre-trained Language Model}
To estimate the relevance score between embeddings of a query and indexed data, it is essential to understand the semantics.
A common method is to build a pre-trained language model as an initial point for a relevance model instead of pre-training from scratch.
Some off-the-shelf language models, such as BERT \cite{devlin2019bert} and RoBERTa \cite{liu2019roberta}, represent promising results in various downstream tasks.
However, they are not suitable for retrieving mobile features since those models were trained with a general corpus without specific domain knowledge that we require.
Thus, we build our own language model as a backbone for the retrieval system. 
More precisely, we built two language models, one in English and the other in Korean, using Transformer encoder \cite{vaswani2017attention}.
To inject domain knowledge to the language models, we construct a training corpus consisting of Wikipedia and smartphone-related articles on the websites such as Samsung Newsroom\footnote{https://news.samsung.com} (an official website introducing Samsung Electronics products), Samsung Members\footnote{https://r1.community.samsung.com} (a community website for users of Samsung products), and the E-manual website\footnote{https://www.samsung.com/mobile} for Galaxy smartphones.  

Moreover, as Korean Wikipedia is much smaller than English Wikipedia, we utilize the newspaper and book corpora released by the National Institute of Korean Language\footnote{https://corpus.korean.go.kr/} as extra training data.

\begin{figure}[t]
\centering
\includegraphics[width=0.9\columnwidth]{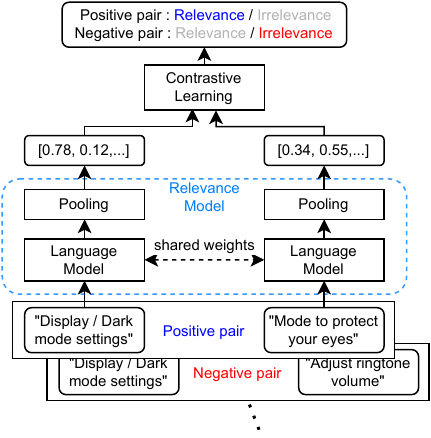} 
\caption{Relevance model with Siamese structures. Each bi-encoder consists of a pre-trained language model and a pooling layer. This model is trained in a contrastive manner with the 1:4 ratio of positive and negative pairs.}
\label{fig2}
\end{figure}

\begin{table}[t]
\centering
\begin{tabular}{cc|cc|cc}
\hline
\multicolumn{2}{c|}{Lang.}                                                                             & \multicolumn{2}{c|}{Eng.}              & \multicolumn{2}{c}{Kor.}              \\ \hline
\multicolumn{2}{c|}{Engine}                                                                            & \multicolumn{1}{c|}{OneUI} & Ours  & \multicolumn{1}{c}{OneUI} & Ours  \\ \hline
\multicolumn{1}{c|}{\multirow{3}{*}{\begin{tabular}[c]{@{}c@{}}Exact\\ keyword\end{tabular}}}        & P  & \multicolumn{1}{c|}{88.7}      & 93.0  & \multicolumn{1}{c|}{74.2}      & 93.2  \\ \cline{2-6} 
\multicolumn{1}{c|}{}                                                                             & R  & \multicolumn{1}{c|}{100.0}     & 100.0 & \multicolumn{1}{c|}{100.0}     & 100.0 \\ \cline{2-6} 
\multicolumn{1}{c|}{}                                                                             & F1 & \multicolumn{1}{c|}{91.2}      & 94.7  & \multicolumn{1}{c|}{80.7}      & 94.8  \\ \hline
\multicolumn{1}{c|}{\multirow{3}{*}{\begin{tabular}[c]{@{}c@{}}Relaxed\\ keyword\end{tabular}}} & P  & \multicolumn{1}{c|}{22.7}      & 36.8  & \multicolumn{1}{c|}{23.6}      & 48.0  \\ \cline{2-6} 
\multicolumn{1}{c|}{}                                                                             & R  & \multicolumn{1}{c|}{35.7}      & 61.9  & \multicolumn{1}{c|}{38.1}      & 72.0  \\ \cline{2-6} 
\multicolumn{1}{c|}{}                                                                             & F1 & \multicolumn{1}{c|}{24.3}      & 42.1  & \multicolumn{1}{c|}{25.5}      & 52.9  \\ 
\hline
\end{tabular}
\caption{Performance on keyword queries. Macro-average values of precision (P), recall (R), and F1 score are reported with top-5 retrievals. For each query, irrelevant retrieved results are regarded as negatives.}
\label{table-key}
\end{table}

\subsection{Relevance Model}
The neural search engine retrieves relevant documents based on the cosine similarity between vector representations.
We need a specialized encoder called a {\em relevance model} to compute a vector representation that captures the relevance in a specific target domain.
It is usual to train the relevance model with query-document pairs  from users in the target search domain.
However, we cannot collect such query-document pairs for mobile features due to privacy issues.
As an alternative, we generated synthetic training pairs by combining the name of each feature and its descriptions instead of actual user queries.
We first extract a feature name and its hint text from the hierarchy in the feature tree.
For example, the  ``\textit{Eye comfort shield}'' feature can be paired with ``\textit{Keep your eyes comfortable by limiting blue light}''.
We also added the descriptions in the product manual, such as ``\textit{Eye comfort shield can help prevent eye strain, especially when you use your phone at night or in low-light settings}.''
In total, the training data consist of 862 English and 911 Korean descriptions of 563 mobile features.

Inspired by \cite{reimers2019sentence}, we built a Siamese network \cite{koch2015siamese} with the pre-trained language model and used  the synthetic pairs to train the relevance model.
As Figure \ref{fig2} illustrates, each output of the language model is average pooled into a single vector.
Then, the relevance score is computed in terms of cosine similarity.
We apply contrastive learning to capture the relevance between a query and documents. 
As for soft negative pairs, we sample irrelevant texts from mini-batch during training as proposed in \cite{Henderson2017EfficientNL}.

\begin{table*}[t]
\centering
\begin{tabular}{|cc|l|}
\hline
\multicolumn{2}{|c|}{Keyword query} & \multicolumn{1}{c|}{Sentence query} \\ \hline
\multicolumn{1}{|c|}{Exact keyword query}  & Relaxed keyword query  & \multirow{2}{*}{\begin{tabular}[c]{@{}l@{}}\textit{``Touch is not working properly.''}\\ \textit{``Screen touch doesn't work when covering screen protector.''}\end{tabular}} \\ \cline{1-2}
\multicolumn{1}{|c|}{\textit{``Touch sensitivity''}}  & \textit{``Mistouch, Block touch''}  &  \\ \hline
\end{tabular}
\caption{Examples of keyword and sentence queries for \textit{``Display - Touch sensitivity''.}}
\label{table-examples}
\end{table*}

\section{Experiments}
Here we demonstrate the effectiveness of our neural search engine on both keyword and sentence queries.
\subsection{Experimental Settings}
For a quantitative evaluation, we compared our engine with OneUI 4.0, the latest version of the customized Settings for Samsung mobile devices based on Android 12, 
which contains 563 mobile features as the search target.
To build the search index, we concatenate the hierarchical path from the root to each node (e.g., \textit{``Display - Touch sensitivity''}).
We acquired all the experimental results in a single run and drew the results with a confidence threshold.
OneUI 4.0 does not provide ranked results since it searches features by matching text in the query and features (Full-Text Search).

We also compare our system with iOS 15.6. Since its menu tree differs from the Android's, a quantitative comparison is not feasible. Instead, we performed the qualitative study using the features common to Android and iOS and examined the first screen of search results, as shown in Figure~\ref{figure-keyword}. Overall, it is clear that the iOS retrieval system also heavily relies on term-matching and cannot do well on semantic search.

\subsection{Evaluation}
To measure the performance of our proposed retrieval system on smartphone features, we collected two kinds of test queries: Keyword queries and sentence queries.
We hired seven annotators who are experts in Android
and asked them to write the keyword and sentence queries for each feature.
In detail, the keyword queries are also categorized into \textit{``exact keyword query''} and  \textit{``relaxed keyword query''}.
The former consists of the exact name of each feature as keywords, while the latter contains alternative keywords describing the feature.
The examples and the test queries' statistics are shown in Tables \ref{table-examples} and \ref{table-statistics}, respectively.

\begin{table}[t]
\centering
\begin{tabular}{|c|c|c|}
\hline
 & English &  Korean \\
\hline
\# Keyword queries in total              & 1,438 & 1,442  \\ \hline
\# Sentence queries in total            & 1,119 & 1,140  \\ \hline
Avg. \# words in a keyword query  & 3.0   & 2.9    \\ \hline 
Avg. \# words in a sentence query       & 6.8   & 4.6    \\
\hline
\end{tabular}
\caption{Statistics of two types of test queries. On average, a sentence query consists of twice the words of a keyword query.}
\label{table-statistics}
\end{table}

\begin{table}[t]
\centering
\begin{tabular}{c|c|c|c|c|c}
\hline
Lang. & Engine & H@5 & H@10 & H@20 & H@all \\
\hline
\multirow{2}{1.5em}{Eng.} & OneUI & 17.5 & 19.2 & 20.0 & 20.4 \\
 & Ours & 76.4 & 83.0 & 89.8 & 100.0 \\
\hline
\multirow{2}{1.5em}{Kor.} & OneUI & 22.0 & 24.7 & 26.2 & 27.4 \\
 & Ours & 83.3 & 89.3 & 93.9 & 100.0 \\
\hline
\end{tabular}
\caption{Performance on sentence queries. Since each sentence query usually has one ground truth, Hits@K is adopted as an evaluation metric}
\label{table-set}
\end{table}

\subsection{Keyword Queries}
We conducted comparative experiments on keyword queries with the currently deployed search system, OneUI 3.1, as a baseline. As shown in Table \ref{table-key}, our system considerably outperforms the baseline in both English and Korean queries.
Moreover, even when users do not know the exact name of the feature (Relaxed keyword query), it shows better performance.
Since the proposed search system utilizes the relevance model, which learns the relevance with a bi-encoder structure, it can retrieve lexically and semantically relevant results.

\subsubsection{Synonym}
Since the keyword-based search in OneUI 3.1 relies on the term frequency, it retrieves disparate results for different queries with similar meanings. However, users may expect the relevant results according to the semantics of the query rather than lexical matching.
With contextualized embeddings of queries and contents, our system can capture the meanings of synonym queries.
As shown in Figure \ref{figure-keyword} (top), our search engine consistently retrieves the desired feature, \textit{``Adapt Sound''}, for the semantically similar query (\textit{``Optimize sound''}) at the ranks 4 and 5 while the baseline misses synonyms and the desired result is ranked at 9. 

\newcommand\wn{2.8}
\newcommand\hn{2.4}
\begin{figure*}[t]
\centering
\subfigure[``\textit{Adapt sound}" : OneUI and ours retrieve corresponding features, but iOS only focuses on lexical matching (e.g. `ad-')]
{
    \put(25, 75){OneUI}
    \frame{\includegraphics[height=\hn cm, width=\wn cm]{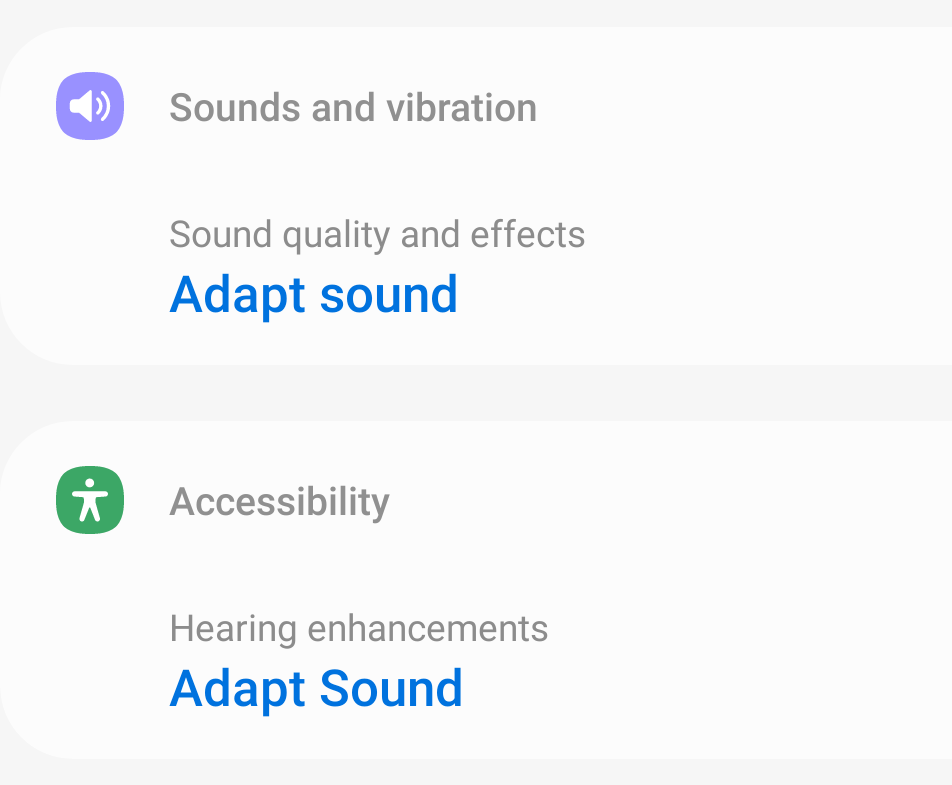}}
    \put(30, 75){Ours}
    \frame{\includegraphics[height=\hn cm, width=\wn cm]{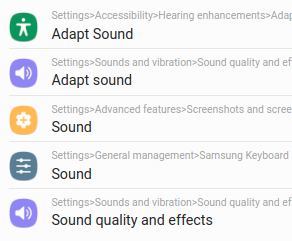}}
    \put(30, 75){iOS}
    \frame{\includegraphics[height=\hn cm, width=\wn cm]{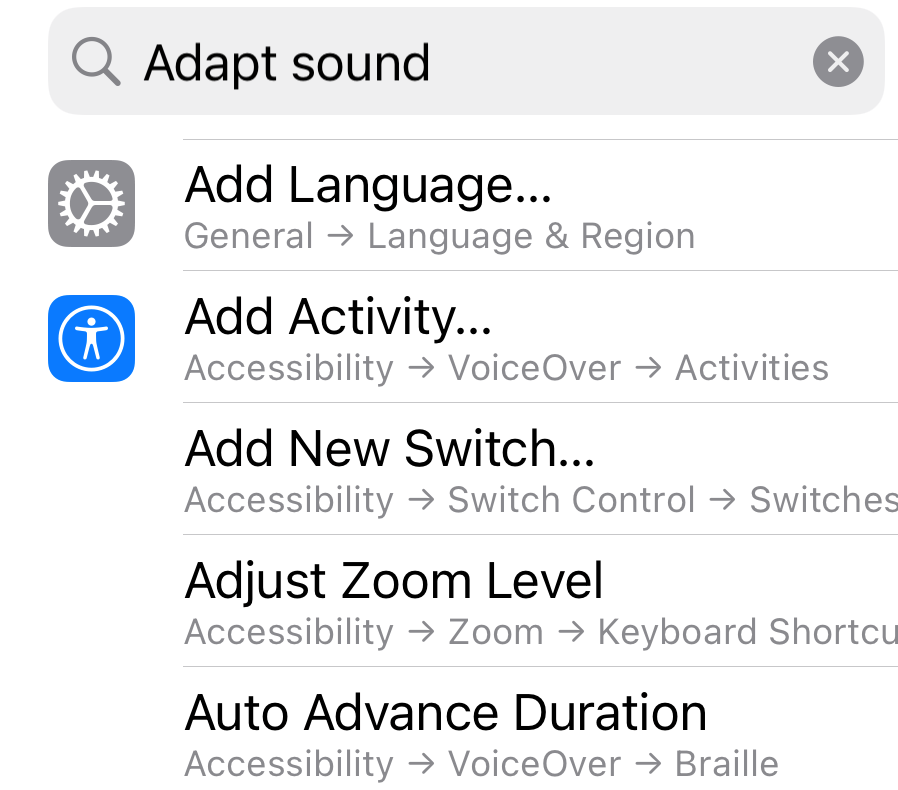}}
}
\hspace{5pt}
\subfigure[``\textit{Optimize sound}" : OneUI shows irrelevant features such as \textit{``Screen recorder - Sound''} and \textit{``Touch feedback - Sound''}.]
{
    \put(25, 75){OneUI}
    \frame{\includegraphics[height=\hn cm, width=\wn cm]{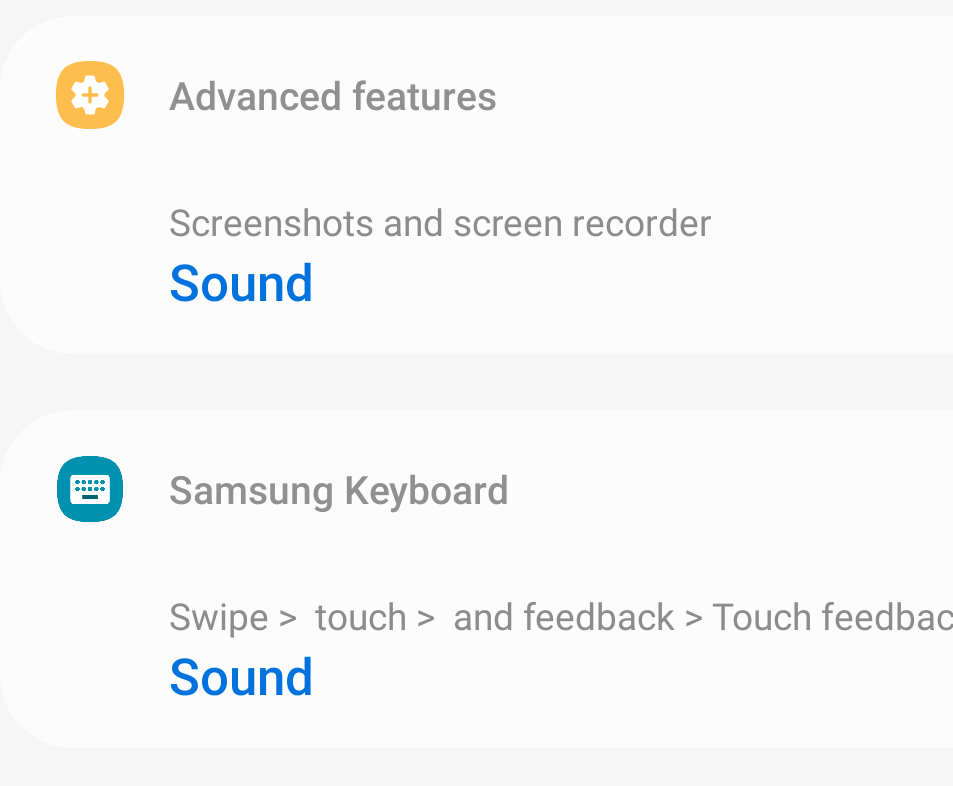}}
    \put(30, 75){Ours}
    \frame{\includegraphics[height=\hn cm, width=\wn cm]{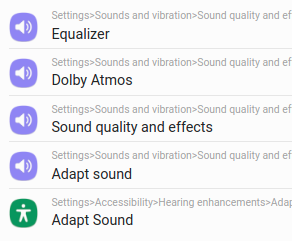}}
    \put(30, 75){iOS}
    \frame{\includegraphics[height=\hn cm, width=\wn cm]{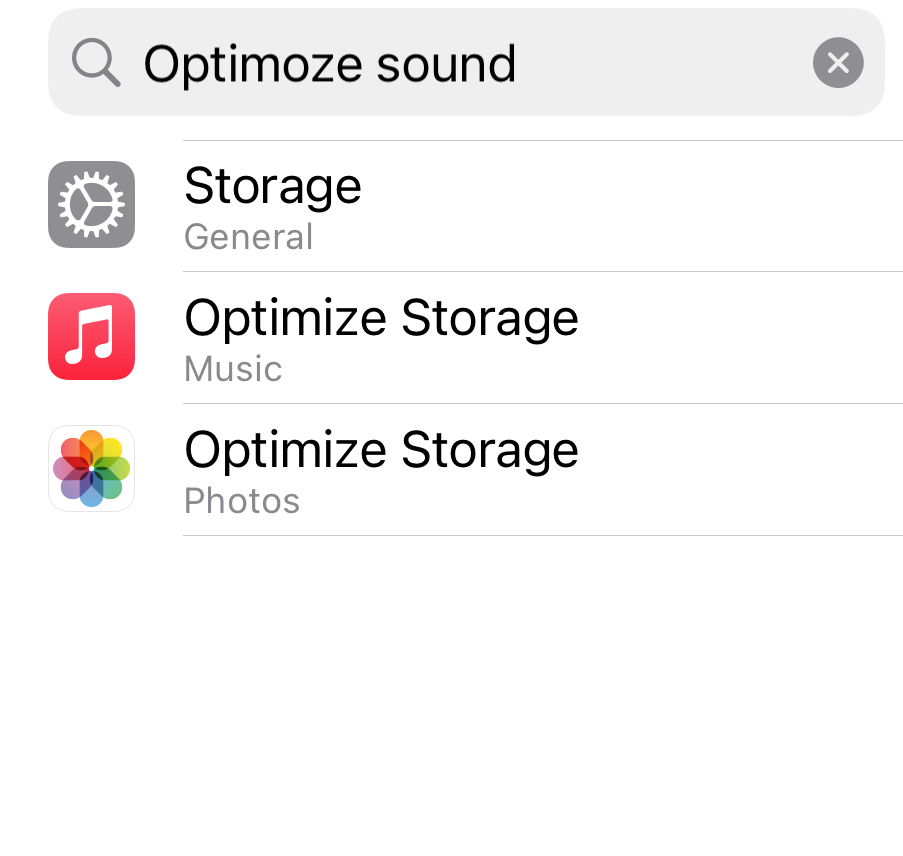}}
}
\subfigure[``\textit{Notification sound}" : all systems retrieve the desired feature \textit{``Notification sound''} in the first rank]
{
    \put(0, 0){}
    \frame{\includegraphics[height=\hn cm, width=\wn cm]{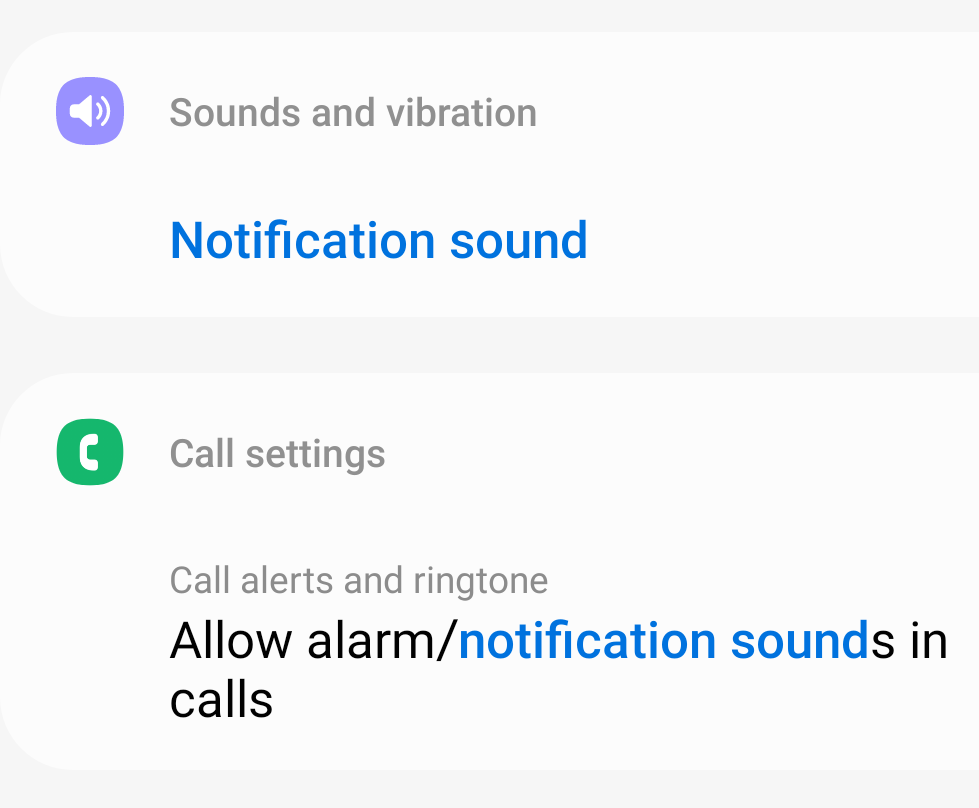}}
    \put(0, 0){}
    \frame{\includegraphics[height=\hn cm, width=\wn cm]{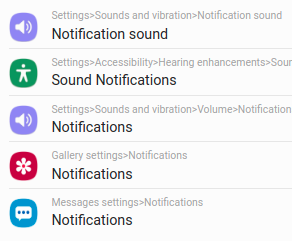}}
    \put(0, 0){}
    \frame{\includegraphics[height=\hn cm, width=\wn cm]{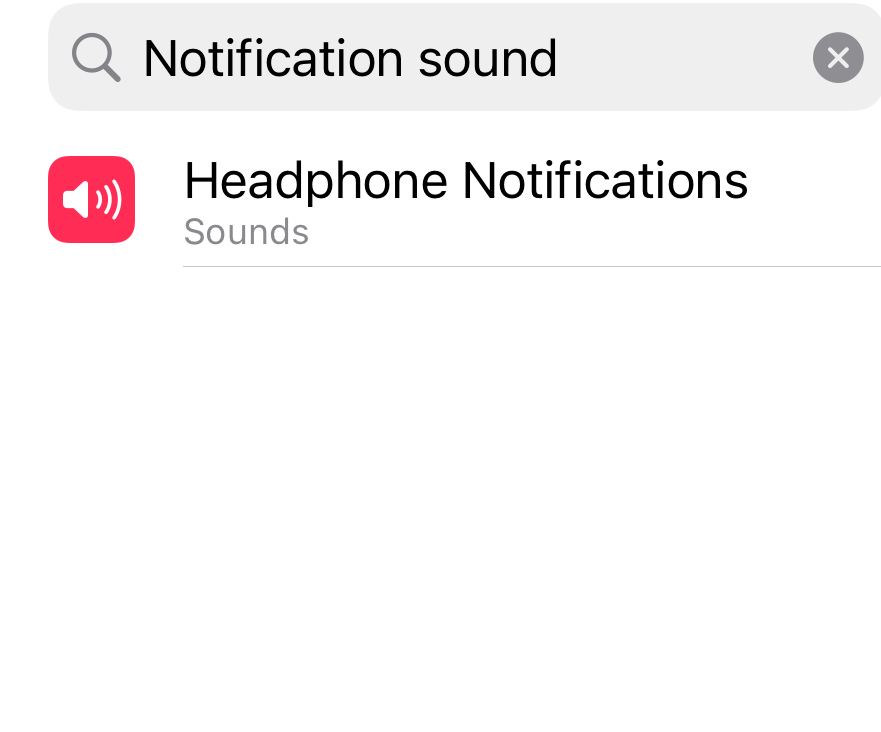}}
}
\hspace{5pt}
\subfigure[``\textit{Sound notification}" : Only ours can distinguish between \textit{``Sound Notifications''} (first) and \textit{``Notification sound''} (second)]
{
    \put(0, 0){}
    \frame{\includegraphics[height=\hn cm, width=\wn cm]{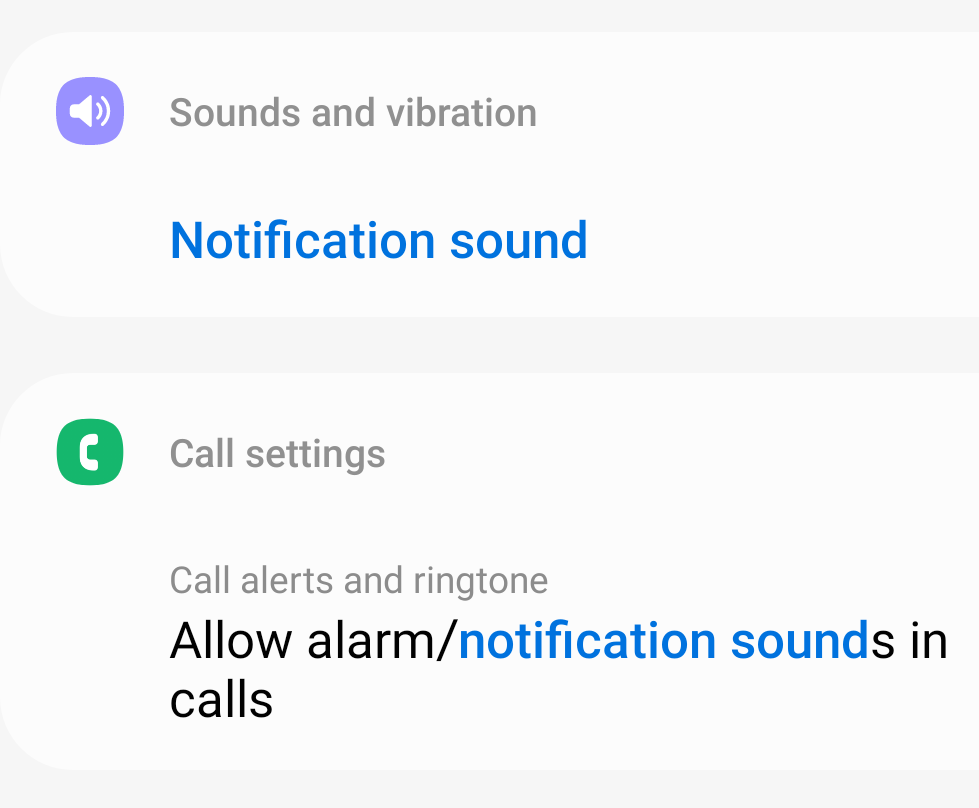}}
    \put(0, 0){}
    \frame{\includegraphics[height=\hn cm, width=\wn cm]{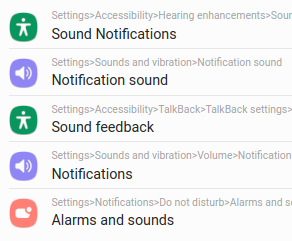}}
    \put(0, 0){}
    \frame{\includegraphics[height=\hn cm, width=\wn cm]{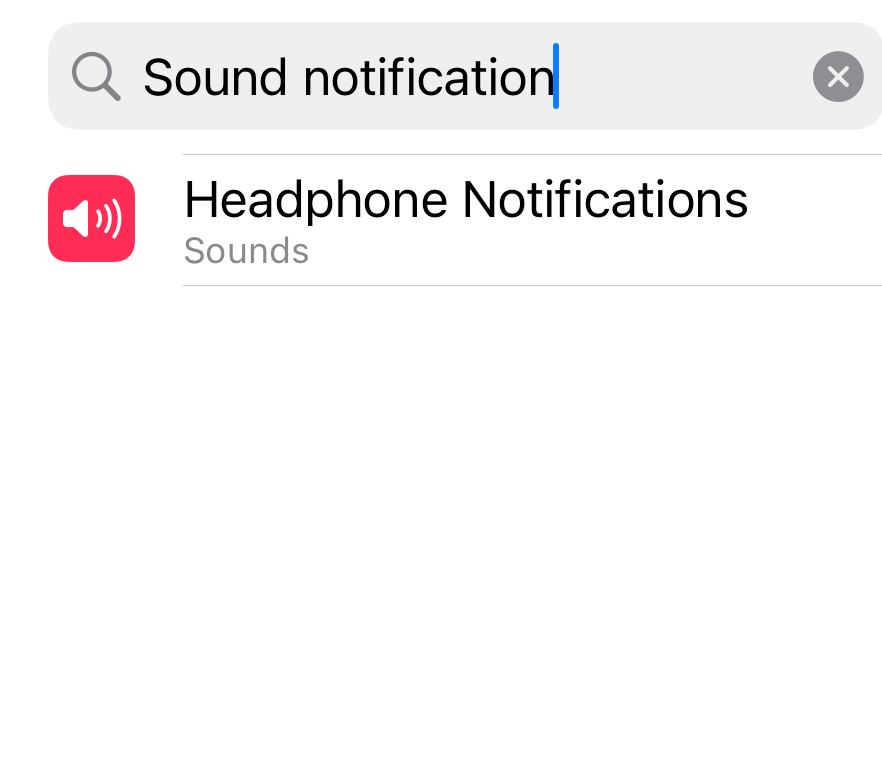}}
}
\caption{Qualitative comparison of keyword queries, including synonym (top) and compound noun (bottom).}
\label{figure-keyword}
\end{figure*}

\subsubsection{Compound nouns}
Permutation of compound nouns is common obstacles for the retriever because they can easily confuse the standard search algorithms, which depend on the exact match.
Since the order of nouns determines the semantic meanings, it is difficult for them to capture what the compound stands for without context information.
For instance, \textit{``Sound notification''} means a type of notification, whereas \textit{``Notification sound''} implies adjusting the volume or style of sound. 
As in Figure \ref{figure-keyword} (bottom), ours ranks the relevant results based on the queries, while the baseline just retrieves naive results, ignoring the order of nouns.

\subsection{Sentence Queries}
We also evaluate the search performance on sentence queries (e.g., \textit{``I want to share internet through my phones''}). This type of queries is also called \textit{interactive queries} since they are often the questions that the user asks to voice assisant systems.
In contrast to keyword queries, each sentence query often has only one ground truth due to its specific intention. Thus, precision and recall are no longer effective in estimating retrieval performance. Instead, we adopt Hits@K as a metric for sentence queries, representing the ratio of the queries which can retrieve at least a single relevant feature in the Top-K results.

\begin{figure}[t]
\centering
\includegraphics[width=0.9\columnwidth]{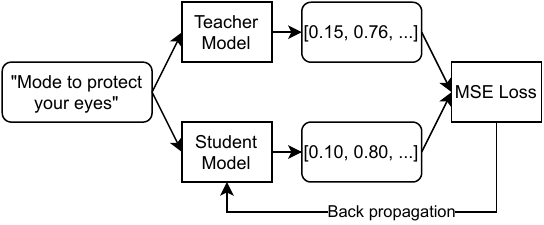}
\caption{The overview of knowledge distillation process. The MSE loss of the output vectors of student and teacher models is used for the back-propagation of the student model.}
\label{fig3}
\end{figure}

As shown in Table \ref{table-set}, OneUI has trouble perceiving the context of sentence queries.
Since term frequency-based approaches rely only on whether the feature includes specific query terms, they are vulnerable to lexically close but semantically different queries.
Notably, the performance on Korean queries is better than the performance on English queries. This gap is attributed to the performance of the underlying language models since we additionally trained the Korean language model with extra corpora.

\begin{figure*}[t]
\centering
\subfigure[\textit{``The letters are too small"} : 
Ours retrieves the features related to font size while others cannot capture the meaning.]
{
    \put(25, 75){OneUI}
    \frame{\includegraphics[height=\hn cm, width=\wn cm]{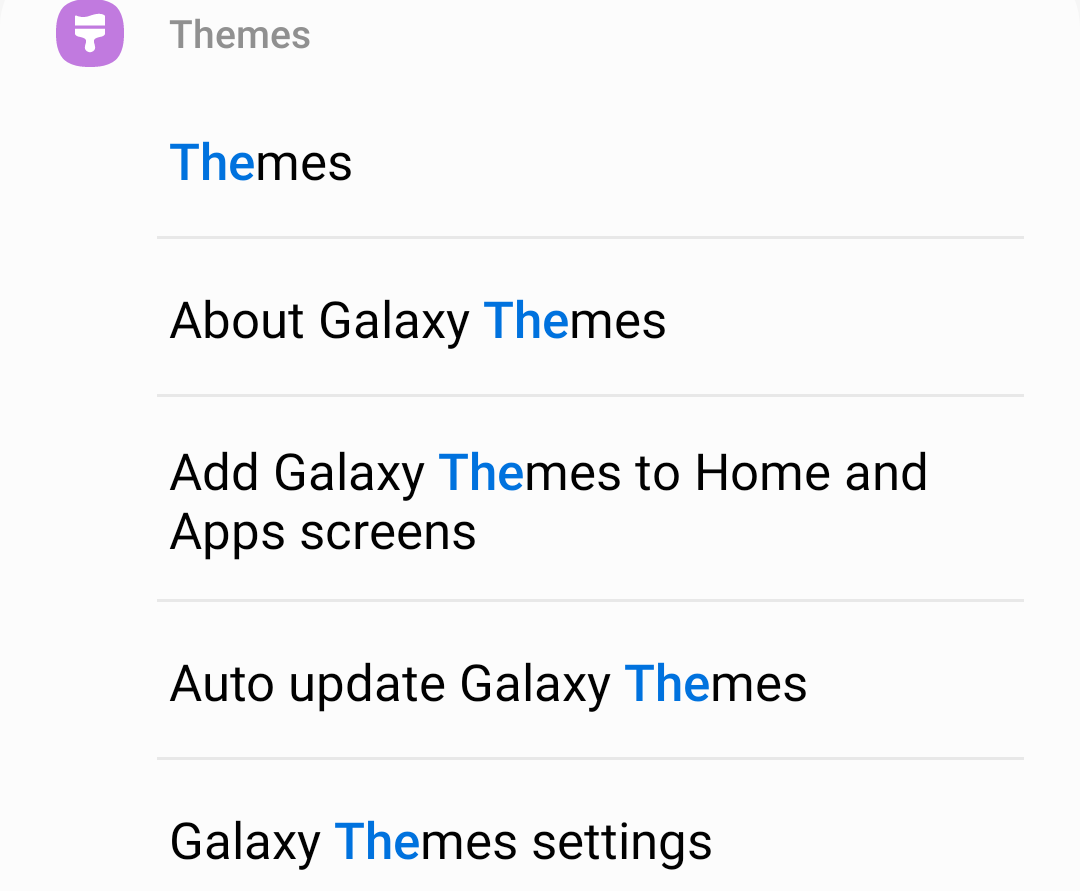}}
    \put(30, 75){Ours}
    \frame{\includegraphics[height=\hn cm, width=\wn cm]{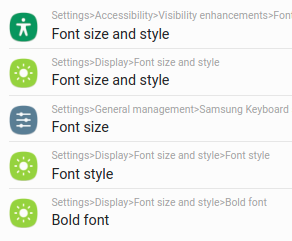}}
    \put(30, 75){iOS}
    \frame{\includegraphics[height=\hn cm, width=\wn cm]{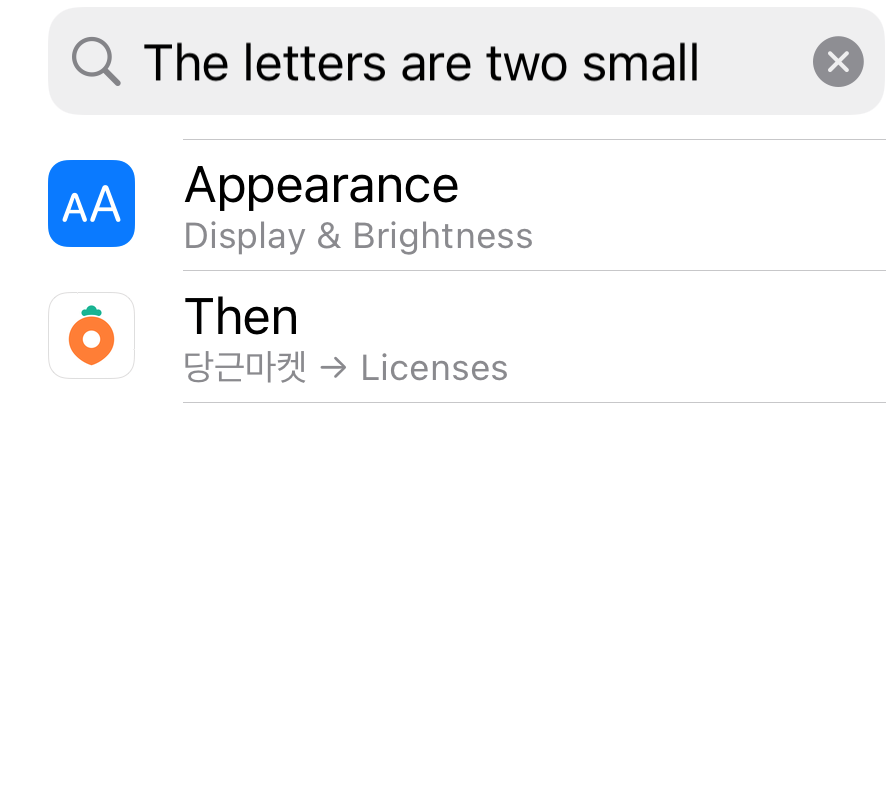}}
}
\hspace{5pt}
\subfigure[\textit{``I wanna share internet through my phone"} : 
Ours retrieves tethering features, but others blindly matches the specific terms.]
{
    \put(25, 75){OneUI}
    \frame{\includegraphics[height=\hn cm, width=\wn cm]{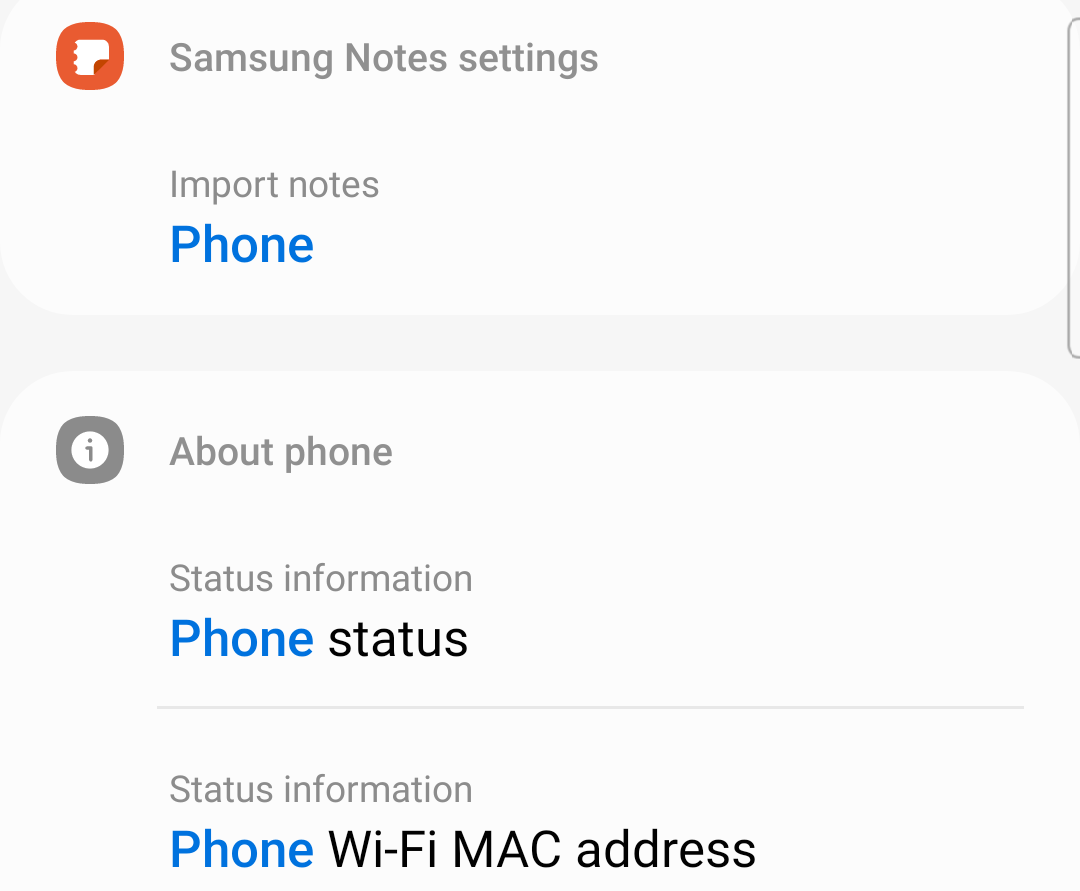}}
    \put(30, 75){Ours}
    \frame{\includegraphics[height=\hn cm, width=\wn cm]{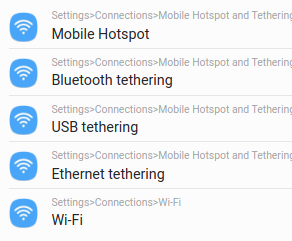}}
    \put(30, 75){iOS}
    \frame{\includegraphics[height=\hn cm, width=\wn cm]{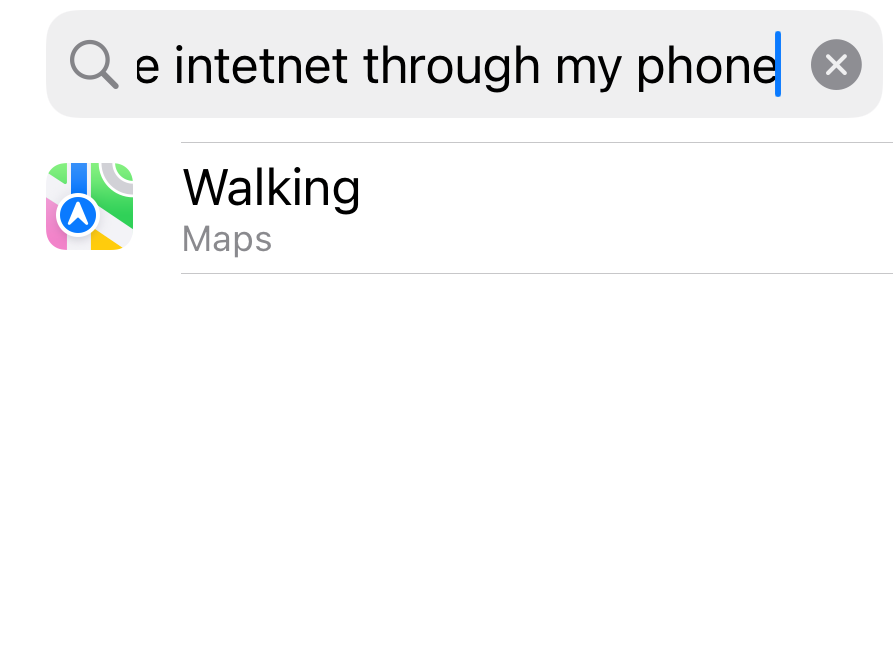}}
}
\subfigure[\textit{``How to dim screen"} : 
Ours retrieves results about screen brightness while others focus on lexical matching]
{
    \put(0, 0){}
    \frame{\includegraphics[height=\hn cm, width=\wn cm]{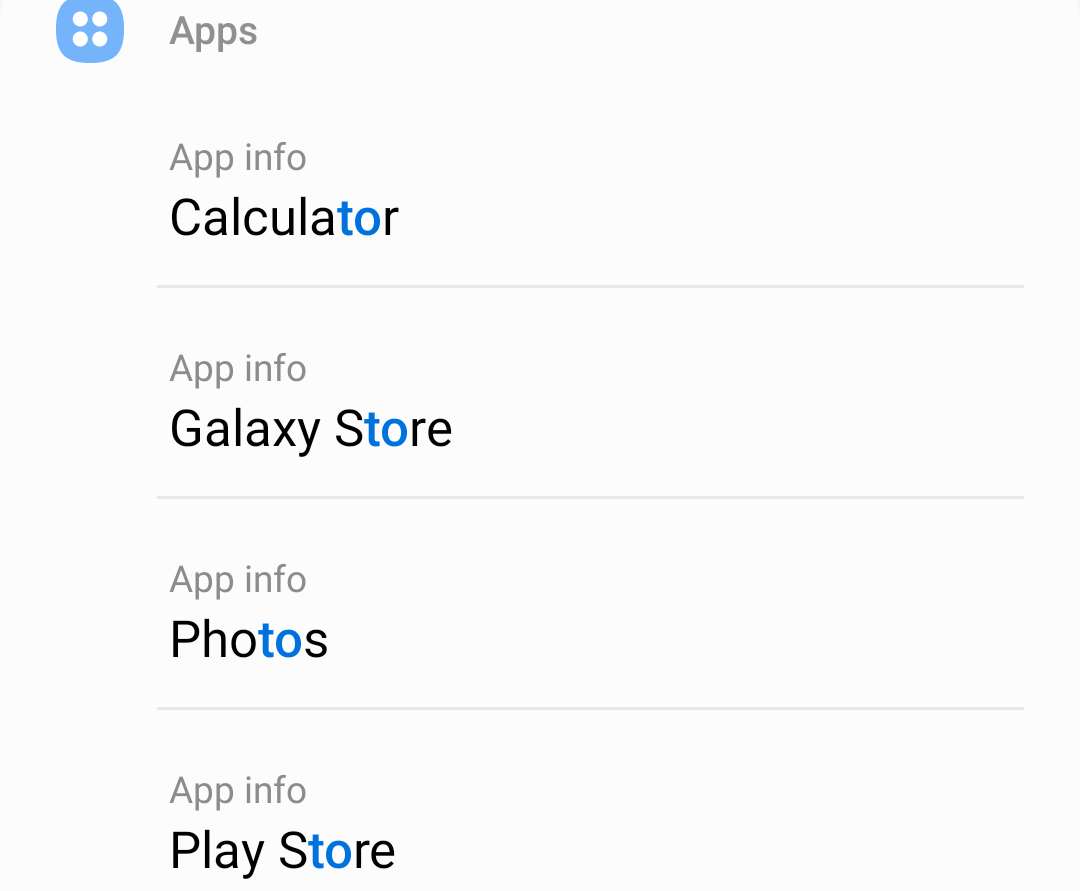}}
    \put(0, 0){}
    \frame{\includegraphics[height=\hn cm, width=\wn cm]{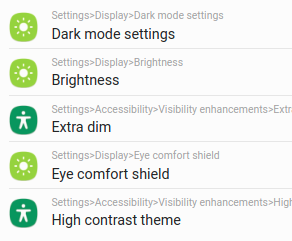}}
    \put(0, 0){}
    \frame{\includegraphics[height=\hn cm, width=\wn cm]{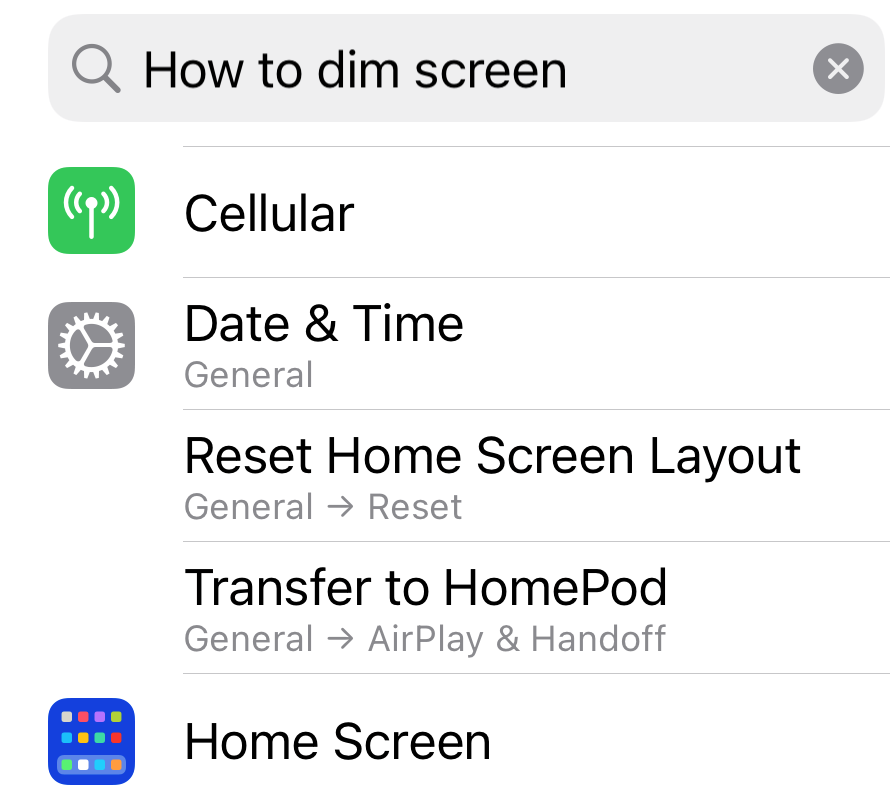}}
}
\hspace{5pt}
\subfigure[\textit{``How to check the remaining battery level"} : 
Only ours retrieves the features related to battery level]
{
    \put(0, 0){}
    \frame{\includegraphics[height=\hn cm, width=\wn cm]{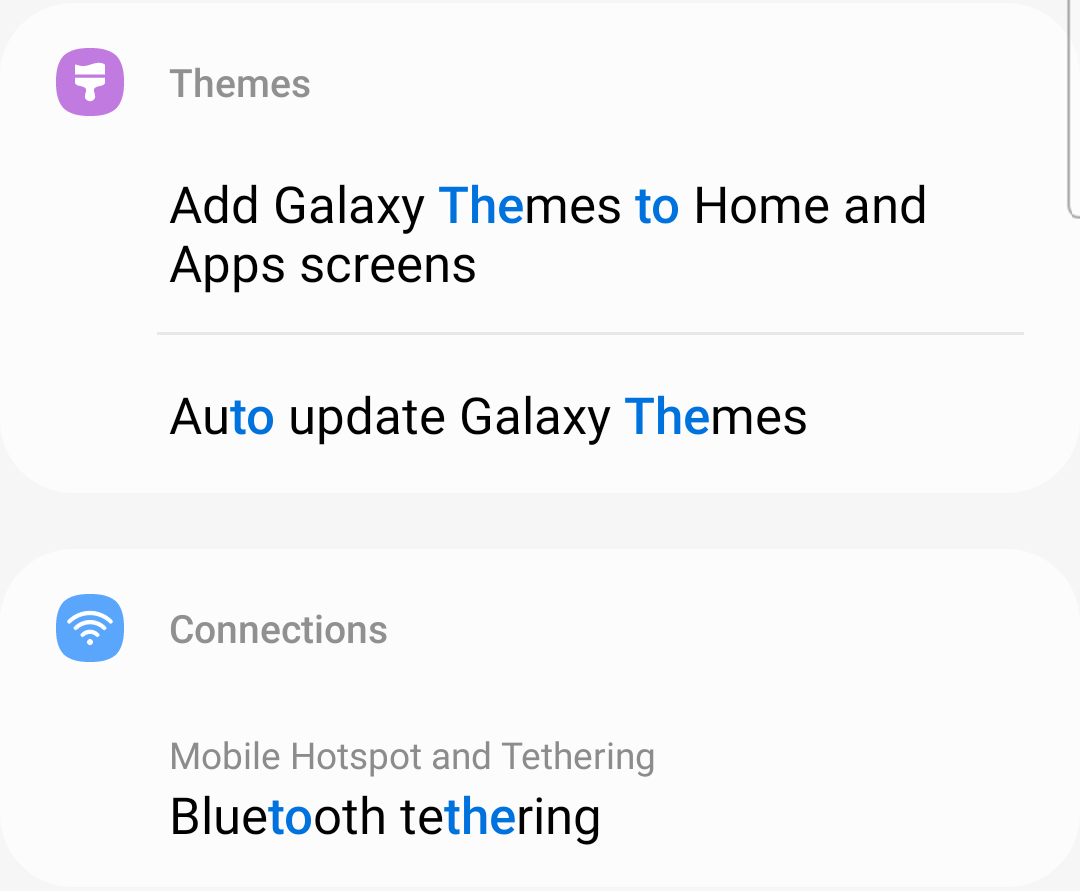}}
    \put(0, 0){}
    \frame{\includegraphics[height=\hn cm, width=\wn cm]{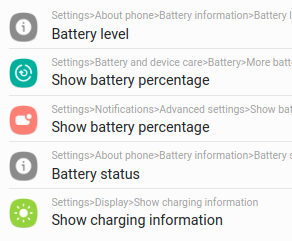}}
    \put(0, 0){}
    \frame{\includegraphics[height=\hn cm, width=\wn cm]{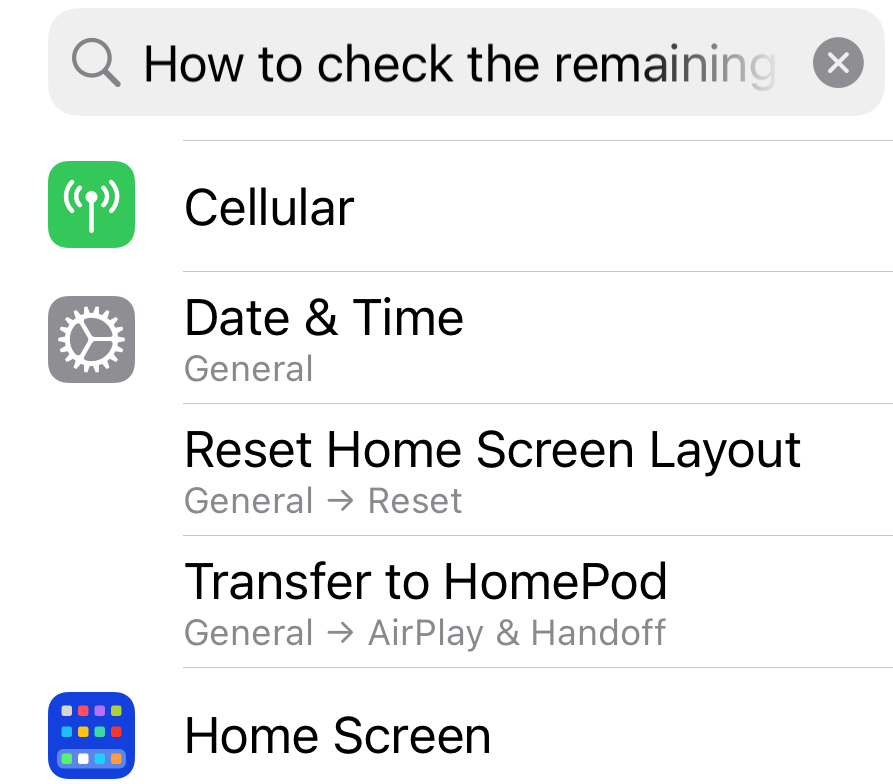}}
}
\caption{Qualitative comparison on sentence queries.}
\label{figure-nl}
\end{figure*}

For qualitative studies, we sample a few sentence queries to verify the retrieving ability of our system. As shown in Figure \ref{figure-nl}, ours can retrieve relevant results such as \textit{``Font size''} and \textit{``Screen brightness''} according to the queries \textit{``The letters are too small''} and \textit{``How to dim screen''}, while the baseline only focused on the specific terms \textit{``the''} and \textit{``to''}, respectively.
In particular, our system is robust to even long queries such as \textit{``I wanna share internet through my phone''} and \textit{``How to check the remaining battery level''}.
Note that the unexpected results of the baseline are not specific to OneUI. For efficiency and simplicity, most search applications adopt the Full-Text Search \cite{bast2013index} as an auxiliary tool that utilizes a look-up table of frequent words. Although this kind of shortcut is advantageous for frequently used queries, it could negatively affect understanding long sentence queries.

\subsection{Knowledge Distillation}
Although retrieving relevant results is the most important, another essential matter for deployment is the size and complexity of the system.
Since mobiles have limited computing resources, it is mandatory to compress the model to minimize response latency and use a small memory.
One of the effective methods to compress a neural model is knowledge distillation \cite{hinton2015distilling}, which transfers the knowledge from the original model (teacher) to a smaller one (student).
In knowledge distillation, the student model is trained to imitate the output vectors or logits of the teacher model. 
Figure \ref{fig3} shows the brief process of knowledge distillation.

The size of our original relevance model is 377MB, which is too large to be deployed on a mobile device. So, we compress it using knowledge distillation.
In particular, we employ the technique for compressing transformer-based encoder models proposed in \cite{turc2019well}.
First, we set a small language model with fewer layers and hidden dimensions as an initial point of the student model.
Then, we train the student model to generate the same embedding with the teacher model.
To this end, we feed entire sentences in the Wikipedia dumps and calculate the Mean Squared Error (MSE) loss between the teacher and student model output embeddings.

\subsubsection{Evaluation}
We applied knowledge distillation to compress our language model with various combinations. We conducted additional experiments on knowledge distillation to investigate the trade-off between distilled models’ size and their performance. 
In transformer-based models, the number of layers (\textsl{L}) and hidden dimensions (\textsl{D}) determine the complexity of the model.
Note that we adopt smaller pre-trained models as the initial state to transfer the learning effect. Table \ref{table-distill} describes the retrieval performance of the distilled models in detail.

\begin{table}[t]
\centering
\begin{tabular}{c|c|c|c|c|c|c}
\hline
\multirow{2}{2.1em}{Lang.} & \multicolumn{3}{c}{Model} & \multicolumn{3}{|c}{Performance} \\
\cline{2-7}
 & \textsl{L} & \textsl{D} & Size & H@5 & H@10 & H@20 \\
\hline
\multirow{4}{2.1em}{Eng.} & 12 & 768 & 377MB & 76.4 & 83.0 & 89.8 \\
\cdashline{2-7}
& 4 & 512 & 82MB & 74.7 & 83.0 & 89.4 \\
& 4 & 256 & 29MB & 72.7 & 80.8 & 87.3 \\
& 2 & 128 & 9.8MB & 67.4 & 76.2 & 84.1 \\
\hline
\multirow{4}{2.1em}{Kor.} & 12 & 768 & 377MB & 83.3 & 89.3 & 93.9 \\
\cdashline{2-7}
& 4 & 512 & 82MB & 82.4 & 88.6 & 93.1 \\
& 4 & 256 & 29MB & 80.1 & 86.9 & 92.2 \\
& 2 & 128 & 9.8MB & 71.9 & 81.0 & 88.1 \\
\hline
\end{tabular}

\caption{Performance of distilled models on sentence queries. \textsl{L} and \textsl{D} denote the number of layers and hidden dimensions. Even after the original model is distilled to student one (up to 2.6\%), it maintains performance.
The original model consists of 12 layers and 768 dimensions.}
\label{table-distill}
\end{table}

The distilled models show competitive performance with moderate sizes compared to the original one.
The distilled English model with 4 layers and 512 hidden dimensions maintained 99.1\% of the performance of the original one in Hits@20 while reducing the model's size to about 20\%. Even in the extreme case when the size is reduced to 2.6\% of the original model, the tiny model still achieves 84.1\% in Hits@20.
Similarly, the distilled Korean one with 4 layers and 512 hidden dimensions maintained 99.1\% of the performance of the original one in Hits@20. In the extreme case, the tiny model which is 2.6\% of the original model still achieves 88.1\% in Hits@20.

\section{Conclusion}
We presented a novel search engine that can retrieve relevant results based on semantic similarity in smartphone settings application.
We built a Siamese architecture with a pre-trained language model and trained the relevance model via contrastive learning with text pairs of the features.
We constructed our own test queries to compare the search performance with the currently-deployed search engine, which adopts Full-Text Search.
The experiments show that the proposed system showed better results than the baseline on both keyword and sentence queries.
Furthermore, we applied the knowledge distillation techniques to compress the models to board on mobile devices achieving reliable search performance.
Clearly, the proposed system is a response to the request to improve the user experience, and we are discussing the deployment plan with the system software group. As the proposed software should be part of the system software (OneUI), rather than a standalone application. The discussion also examines various other issues not directly related to the presented system, such as how to support legacy search results with the neural search and the UI to show the results, etc.
We expect the proposed system to be applied to improve settings search, app store search, and troubleshooting search, where users know the various descriptive and semantic information rather than exact terms.
\bibliography{reference}

\begin{thebibliography}{30}
\providecommand{\natexlab}[1]{#1}

\bibitem[{Anil et~al.(2018)Anil, Pereyra, Passos, Ormandi, Dahl, and
  Hinton}]{anil2018large}
Anil, R.; Pereyra, G.; Passos, A.; Ormandi, R.; Dahl, G.~E.; and Hinton, G.~E.
  2018.
\newblock Large scale distributed neural network training through online
  distillation.
\newblock \emph{arXiv preprint arXiv:1804.03235}.

\bibitem[{Bast and Buchhold(2013)}]{bast2013index}
Bast, H.; and Buchhold, B. 2013.
\newblock An index for efficient semantic full-text search.
\newblock In \emph{Proceedings of the 22nd ACM international conference on
  Information \& Knowledge Management}, 369--378.

\bibitem[{Clark et~al.(2020)Clark, Luong, Le, and Manning}]{clark2020electra}
Clark, K.; Luong, M.-T.; Le, Q.~V.; and Manning, C.~D. 2020.
\newblock Electra: Pre-training text encoders as discriminators rather than
  generators.
\newblock \emph{arXiv preprint arXiv:2003.10555}.

\bibitem[{Dehghani et~al.(2017)Dehghani, Zamani, Severyn, Kamps, and
  Croft}]{10.1145/3077136.3080832}
Dehghani, M.; Zamani, H.; Severyn, A.; Kamps, J.; and Croft, W.~B. 2017.
\newblock Neural Ranking Models with Weak Supervision.
\newblock In \emph{Proceedings of the 40th International ACM SIGIR Conference
  on Research and Development in Information Retrieval}, SIGIR '17, 65–74.
  New York, NY, USA: Association for Computing Machinery.
\newblock ISBN 9781450350228.

\bibitem[{Devlin et~al.(2019)Devlin, Chang, Lee, and
  Toutanova}]{devlin2019bert}
Devlin, J.; Chang, M.-W.; Lee, K.; and Toutanova, K. 2019.
\newblock BERT: Pre-training of Deep Bidirectional Transformers for Language
  Understanding.
\newblock In \emph{NAACL-HLT (1)}.

\bibitem[{Dinan et~al.(2018)Dinan, Roller, Shuster, Fan, Auli, and
  Weston}]{dinan2018wizard}
Dinan, E.; Roller, S.; Shuster, K.; Fan, A.; Auli, M.; and Weston, J. 2018.
\newblock Wizard of wikipedia: Knowledge-powered conversational agents.
\newblock \emph{arXiv preprint arXiv:1811.01241}.

\bibitem[{Henderson et~al.(2017)Henderson, Al-Rfou, Strope, Sung, Luk{\'a}cs,
  Guo, Kumar, Miklos, and Kurzweil}]{Henderson2017EfficientNL}
Henderson, M.; Al-Rfou, R.; Strope, B.; Sung, Y.-H.; Luk{\'a}cs, L.; Guo, R.;
  Kumar, S.; Miklos, B.; and Kurzweil, R. 2017.
\newblock Efficient Natural Language Response Suggestion for Smart Reply.
\newblock \emph{ArXiv}, abs/1705.00652.

\bibitem[{Hinton, Vinyals, and Dean(2015)}]{hinton2015distilling}
Hinton, G.; Vinyals, O.; and Dean, J. 2015.
\newblock Distilling the knowledge in a neural network.
\newblock \emph{arXiv preprint arXiv:1503.02531}.

\bibitem[{Horita, J{\'u}nior, and J{\'u}nior(2019)}]{horita2019enhancing}
Horita, L.~R.; J{\'u}nior, J. B. P.~M.; and J{\'u}nior, M. D.~P. 2019.
\newblock Enhancing the Search Tool of the Android Settings through Natural
  Language Processing.
\newblock In \emph{Anais Estendidos do IX Simp{\'o}sio Brasileiro de Engenharia
  de Sistemas Computacionais}, 83--88. SBC.

\bibitem[{Jones(1972)}]{jones1972statistical}
Jones, K.~S. 1972.
\newblock A statistical interpretation of term specificity and its application
  in retrieval.
\newblock \emph{Journal of documentation}.

\bibitem[{Kim and Rush(2016)}]{kim2016sequence}
Kim, Y.; and Rush, A.~M. 2016.
\newblock Sequence-level knowledge distillation.
\newblock \emph{arXiv preprint arXiv:1606.07947}.

\bibitem[{Koch et~al.(2015)Koch, Zemel, Salakhutdinov et~al.}]{koch2015siamese}
Koch, G.; Zemel, R.; Salakhutdinov, R.; et~al. 2015.
\newblock Siamese neural networks for one-shot image recognition.
\newblock In \emph{ICML deep learning workshop}, volume~2. Lille.

\bibitem[{Liu et~al.(2019)Liu, Ott, Goyal, Du, Joshi, Chen, Levy, Lewis,
  Zettlemoyer, and Stoyanov}]{liu2019roberta}
Liu, Y.; Ott, M.; Goyal, N.; Du, J.; Joshi, M.; Chen, D.; Levy, O.; Lewis, M.;
  Zettlemoyer, L.; and Stoyanov, V. 2019.
\newblock RoBERTa: A Robustly Optimized BERT Pretraining Approach.
\newblock arXiv:1907.11692.

\bibitem[{Luhn(1957)}]{luhn1957statistical}
Luhn, H.~P. 1957.
\newblock A statistical approach to mechanized encoding and searching of
  literary information.
\newblock \emph{IBM Journal of research and development}, 1(4): 309--317.

\bibitem[{Mazar{\'e} et~al.(2018)Mazar{\'e}, Humeau, Raison, and
  Bordes}]{mazare2018training}
Mazar{\'e}, P.-E.; Humeau, S.; Raison, M.; and Bordes, A. 2018.
\newblock Training millions of personalized dialogue agents.
\newblock \emph{arXiv preprint arXiv:1809.01984}.

\bibitem[{Mikolov et~al.(2013)Mikolov, Sutskever, Chen, Corrado, and
  Dean}]{10.5555/2999792.2999959}
Mikolov, T.; Sutskever, I.; Chen, K.; Corrado, G.; and Dean, J. 2013.
\newblock Distributed Representations of Words and Phrases and Their
  Compositionality.
\newblock In \emph{Proceedings of the 26th International Conference on Neural
  Information Processing Systems - Volume 2}, NIPS'13, 3111–3119. Red Hook,
  NY, USA: Curran Associates Inc.

\bibitem[{Mitra and Craswell(2017)}]{DBLP:journals/corr/MitraC17}
Mitra, B.; and Craswell, N. 2017.
\newblock Neural Models for Information Retrieval.
\newblock \emph{CoRR}, abs/1705.01509.

\bibitem[{Mitra and Craswell(2018)}]{Mitra2018AnIT}
Mitra, B.; and Craswell, N. 2018.
\newblock An Introduction to Neural Information Retrieval.
\newblock \emph{Found. Trends Inf. Retr.}, 13: 1--126.

\bibitem[{Mitra, Diaz, and Craswell(2017)}]{10.1145/3038912.3052579}
Mitra, B.; Diaz, F.; and Craswell, N. 2017.
\newblock Learning to Match Using Local and Distributed Representations of Text
  for Web Search.
\newblock In \emph{Proceedings of the 26th International Conference on World
  Wide Web}, WWW '17, 1291–1299. Republic and Canton of Geneva, CHE:
  International World Wide Web Conferences Steering Committee.
\newblock ISBN 9781450349130.

\bibitem[{Park et~al.(2019)Park, Kim, Lu, and Cho}]{park2019relational}
Park, W.; Kim, D.; Lu, Y.; and Cho, M. 2019.
\newblock Relational knowledge distillation.
\newblock In \emph{Proceedings of the IEEE/CVF Conference on Computer Vision
  and Pattern Recognition}, 3967--3976.

\bibitem[{Phuong and Lampert(2019)}]{phuong2019towards}
Phuong, M.; and Lampert, C. 2019.
\newblock Towards understanding knowledge distillation.
\newblock In \emph{International Conference on Machine Learning}, 5142--5151.
  PMLR.

\bibitem[{Reimers and Gurevych(2019)}]{reimers2019sentence}
Reimers, N.; and Gurevych, I. 2019.
\newblock Sentence-BERT: Sentence Embeddings using Siamese BERT-Networks.
\newblock In \emph{Proceedings of the 2019 Conference on Empirical Methods in
  Natural Language Processing and the 9th International Joint Conference on
  Natural Language Processing (EMNLP-IJCNLP)}, 3982--3992.

\bibitem[{Robertson et~al.(1995)Robertson, Walker, Jones, Hancock-Beaulieu,
  Gatford et~al.}]{robertson1995okapi}
Robertson, S.~E.; Walker, S.; Jones, S.; Hancock-Beaulieu, M.~M.; Gatford, M.;
  et~al. 1995.
\newblock Okapi at TREC-3.
\newblock \emph{Nist Special Publication Sp}, 109: 109.

\bibitem[{Salton and McGill(1986)}]{salton1986introduction}
Salton, G.; and McGill, M.~J. 1986.
\newblock \emph{Introduction to Modern Information Retrieval}.
\newblock USA: McGraw-Hill, Inc.
\newblock ISBN 0070544840.

\bibitem[{Turc et~al.(2019)Turc, Chang, Lee, and Toutanova}]{turc2019well}
Turc, I.; Chang, M.-W.; Lee, K.; and Toutanova, K. 2019.
\newblock Well-read students learn better: On the importance of pre-training
  compact models.
\newblock \emph{arXiv preprint arXiv:1908.08962}.

\bibitem[{Vaswani et~al.(2017)Vaswani, Shazeer, Parmar, Uszkoreit, Jones,
  Gomez, Kaiser, and Polosukhin}]{vaswani2017attention}
Vaswani, A.; Shazeer, N.; Parmar, N.; Uszkoreit, J.; Jones, L.; Gomez, A.~N.;
  Kaiser, {\L}.; and Polosukhin, I. 2017.
\newblock Attention is all you need.
\newblock In \emph{Advances in neural information processing systems},
  5998--6008.

\bibitem[{Vig and Ramea(2019)}]{vig2019comparison}
Vig, J.; and Ramea, K. 2019.
\newblock Comparison of transfer-learning approaches for response selection in
  multi-turn conversations.
\newblock In \emph{Workshop on DSTC7}.

\bibitem[{Wolf et~al.(2019)Wolf, Sanh, Chaumond, and
  Delangue}]{wolf2019transfertransfo}
Wolf, T.; Sanh, V.; Chaumond, J.; and Delangue, C. 2019.
\newblock Transfertransfo: A transfer learning approach for neural network
  based conversational agents.
\newblock \emph{arXiv preprint arXiv:1901.08149}.

\bibitem[{Xiong et~al.(2017)Xiong, Dai, Callan, Liu, and
  Power}]{10.1145/3077136.3080809}
Xiong, C.; Dai, Z.; Callan, J.; Liu, Z.; and Power, R. 2017.
\newblock End-to-End Neural Ad-Hoc Ranking with Kernel Pooling.
\newblock In \emph{Proceedings of the 40th International ACM SIGIR Conference
  on Research and Development in Information Retrieval}, SIGIR '17, 55–64.
  New York, NY, USA: Association for Computing Machinery.
\newblock ISBN 9781450350228.

\bibitem[{Zhang et~al.(2019)Zhang, Song, Gao, Chen, Bao, and
  Ma}]{zhang2019your}
Zhang, L.; Song, J.; Gao, A.; Chen, J.; Bao, C.; and Ma, K. 2019.
\newblock Be your own teacher: Improve the performance of convolutional neural
  networks via self distillation.
\newblock In \emph{Proceedings of the IEEE/CVF International Conference on
  Computer Vision}, 3713--3722.

\end{thebibliography}
\end{document}